\documentclass[article]{aastex631}

\usepackage[utf8]{inputenc}

\usepackage{booktabs}
\usepackage{newtxtext,newtxmath}
\usepackage{ulem}
\usepackage{amssymb}
\usepackage{graphicx}

\begin{document}

\title{Quasars: standard candles up to z=7.5 with the precision of Supernovae Ia}

\author{M. G. Dainotti}
\altaffiliation{First and second authors share the same contribution}
\altaffiliation{Corresponding author. E-mail: mariagiovannadainotti@yahoo.it, maria.dainotti@nao.ac.jp}
\affiliation{National Astronomical Observatory of Japan, 2 Chome-21-1 Osawa, Mitaka, Tokyo 181-8588, Japan}
\affiliation{The Graduate University for Advanced Studies, SOKENDAI, Shonankokusaimura, Hayama, Miura District, Kanagawa 240-0193, Japan}
\affiliation{Space Science Institute, 4765 Walnut St, Suite B, 80301 Boulder, CO, USA}

\author{G. Bargiacchi}
\affiliation{Scuola Superiore Meridionale, Largo S. Marcellino 10, 80138 Napoli, Italy}
\affiliation{Istituto Nazionale di Fisica Nucleare (INFN), Sez. di Napoli, Complesso Univ. Monte S. Angelo, Via Cinthia 9, 80126, Napoli, Italy}

\author{A. Ł. Lenart}
\affiliation{Astronomical Observatory, Jagiellonian University, ul. Orla 171, 31-501 Kraków, Poland}

\author{S. Nagataki}
\affiliation{Interdisciplinary Theoretical \& Mathematical Science Program, RIKEN (iTHEMS), 2-1 Hirosawa, Wako, Saitama, Japan 351-0198}
\affiliation{RIKEN Cluster for Pioneering Research, Astrophysical Big Bang Laboratory (ABBL), 2-1 Hirosawa, Wako, Saitama, Japan 351-0198}
\affiliation{Astrophysical Big Bang Group (ABBG), Okinawa Institute of Science and Technology Graduate University (OIST),
1919-1 Tancha, Okinawa, Japan 904-0495}

\author{S. Capozziello}
\affiliation{Scuola Superiore Meridionale, Largo S. Marcellino 10, 80138 Napoli, Italy}
\affiliation{Istituto Nazionale di Fisica Nucleare (INFN), Sez. di Napoli, Complesso Univ. Monte S. Angelo, Via Cinthia 9, 80126, Napoli, Italy}
\affiliation{Dipartimento di Fisica "E. Pancini", Università degli Studi di Napoli Federico II, Complesso Univ. Monte S. Angelo, Via Cinthia 9, 80126, Napoli, Italy}



\begin{abstract}

Currently, the $\Lambda$ Cold Dark Matter model, which relies on the existence of cold dark matter and a cosmological constant $\Lambda$, best describes the Universe. However, we lack information in the high-redshift ($z$) region between Type Ia Supernovae (SNe Ia) (up to $z=2.26$) and the Cosmic Microwave Background ($z=1100$), an interval crucial to test cosmological models and their possible evolution. We have defined a sample of 983 Quasars up to $z=7.54$ with reduced intrinsic dispersion $\delta=0.007$ which determines the matter density parameter $\Omega_M$ with the same precision of SNe Ia. Although previous analysis have been used Quasars as cosmological tools (e.g. Risaliti and Lusso 2019), this is the first time that high-redshift sources, in this case Quasars, as standalone cosmological probes yield such tight constraints on $\Omega_M$. Our results show the importance of correcting cosmological relationships for selection biases and redshift evolution and how the choice of a golden sample reduces considerably the intrinsic scatter. This proves the reliability of Quasars as standard cosmological candles.

\end{abstract}



\section{Introduction} \label{sec:intro}

Nowadays it is time for the scavenger hunt for cosmological parameters such as the Hubble constant, $H_0$, and the matter density parameter, $\Omega_M$.
The main challenge is to further cast light on which is the most suitable description of the Universe.
Currently, the most accredited model is the $\Lambda$CDM model, which includes a cold dark matter (CDM) component and a dark energy associated with a cosmological constant $\Lambda$, as suggested by the current accelerated expansion of the Universe \citep{riess1998,perlmutter1999}. Predictions from this scenario are compatible with observational probes such as the cosmic microwave background (CMB) \citep{planck2018} at redshift $z=1100$, the baryon acoustic oscillations \citep{eboss2021}, and type Ia supernovae (SNe Ia) observed at low $z$ up to $z=2.26$ \citep{Rodney}.
However, we lack information on intermediate redshifts between SNe Ia and CMB.
Having reliable cosmological candles at intermediate redshifts between the SNe Ia and CMB could be the crucial turning point to understanding to what extent the current cosmological $\Lambda$CDM model is the most suitable one. 
To this end, the community strives to reduce as much as possible the uncertainties on cosmological parameters and to find probes at intermediate redshifts which can be considered standard candles \citep{Yonetoku_2004}. 
Given the current interest of the community in Quasars as reliable cosmological tools \citep{DainottiQSO2022,LenartbiasfreeQSO2022,rl19,Bargiacchi2022MNRAS.515.1795B,Khadka2022arXiv}, we focus on the hunt for a gold sample for Quasars, as already done for Gamma-ray Bursts (GRBs) \citep{Dainotti2010ApJ...722L.215D, Dainotti2013a,dainotti2015b,Dainotti2016ApJ...825L..20D, Dainotti2020a,Dainotti2022ApJS..261...25D,Dainotti2021PASJ...73..970D,Dainotti2022PASJ,Dainotti2022MNRAS.514.1828D,Dainotti2023MNRAS.518.2201D}.
We propose an approach to the use of Quasars as standard candles similar to the one done for SNe Ia 25 years ago, but Quasars are observed at much larger distances (up to $z=7.54$).
To tackle such an approach we need a standardizable candle and in this regard, a relation exists between the X-ray and the Ultraviolet (UV) luminosities of Quasars  (known as the Risaliti-Lusso relation, hereafter called RL). To use this, Quasar emission mechanisms need to be very well understood and the relation at play should not be affected by redshift evolution (a change in redshift) and selection biases. 
Namely, such a relation should remain the same at all redshifts or if there is a redshift evolution this should be properly accounted for before its use as a cosmological tool.
\cite{DainottiQSO2022} have already demonstrated that the RL relation is not induced by selection biases or redshift evolution, but it is intrinsic to the physics of Quasars.
Here we have presented two main gold samples of Quasars. One is built from a flat $\Lambda$CDM model with $\Omega_M=0.3$ and  $H_0 = 70  \, \mathrm{km} \, \mathrm{s}^{-1} \, \mathrm{Mpc}^{-1}$ and is composed of 983 Quasars up to $z=7.54$ which constrains the parameter $\Omega_M$ in the assumed cosmology with the same precision as SNe Ia in the Pantheon sample. The other one is instead obtained through a circularity-free procedure and consists of 975 sources with the same maximum redshift of $z=7.54$. 
The paper is structured in data and methodology, Sect. \ref{dataset}, where we detail the sample and all methods, results, Sect. \ref{results} and \ref{MCMC}, where the choice of the golden sample assuming a given cosmology is explained and the solution for a golden sample completely independent from the circularity problem including the Markov Chain sampling uncertainty results is presented. We discuss our findings in Sec. \ref{discussions}. 

\section{Data set and Methodology}\label{dataset}
Our initial Quasar sample is the most recent one released for cosmological studies \citep{Lusso2020A&A...642A.150L}. It consists of 2421 sources in the redshift range between $z=0.009$ and $z =7.54$ \citep{banados2018} collected from eight catalogues \citep{2019A&A...632A.109N,salvestrini2019,2019A&A...630A.118V} and archives \citep{2016yCat..74570110M,2018A&A...613A..51P,2020A&A...641A.136W,2010ApJS..189...37E}, with the addition of a subsample of low redshift Quasars that present UV observations from the International Ultraviolet Explorer and X-ray data in archives. 
To obtain this Quasar sample suitable for cosmological analyses, as many as possible observational biases have been meticulously inspected and removed \citep{rl15,lr16,rl19,salvestrini2019,Lusso2020A&A...642A.150L}. 
We note here some differences between our sample and the samples used in previous works.
Here, we study this final sample of 2421 sources without any additional selection, such as the cut at redshift $z=0.7$ previously used in some works \citep{Lusso2020A&A...642A.150L,Bargiacchi2022MNRAS.515.1795B}, to avoid any possible induced bias due to the reduction on the redshift sample \citep{DainottiQSO2022,LenartbiasfreeQSO2022}.

\subsection{Correction for the redshift evolution of the luminosities}
We clarify that previous works have not considered selection biases and redshift evolution with the exception of our works in the literatures \citep{DainottiQSO2022,LenartbiasfreeQSO2022,Bargiacchi2023MNRAS.tmp..761B}.
Since Quasars are high-redshift sources, we need to account for selection biases and redshift evolution effects. These factors could induce artificial correlations between intrinsic physical quantities of sources \citep{Dainotti2013a}, such as the relation between X-ray and UV luminosities for Quasars. To correct for these effects, we have applied the statistical Efron \& Petrosian method \citep{1992ApJ...399..345E} assuming that the luminosities evolve with redshift as $(1+z)^k$. This method has already been employed for GRBs \citep{Dainotti2013a,dainotti2015b,Dainotti2017A&A...600A..98D,dainotti2021Galax...9...95D,2023MNRAS.518.2201D} and Quasars \citep{DainottiQSO2022,LenartbiasfreeQSO2022}. The choice of a more complex function for the redshift evolution would not affect the results \citep{2011ApJ...743..104S,Dainotti2021ApJ...914L..40D,DainottiQSO2022}. Thus, the de-evolved UV and X-ray luminosities are computed as $L'_{UV}= L_{UV}/(1+z)^{k_{UV}}$ and $L'_{X}= L_{X}/(1+z)^{k_{X}}$ by using $L_{UV}$ and $L_X$ obtained from the measured flux densities $F_{UV}$ and $F_X$ (in units of $\mathrm{erg \, s^{-1} \, cm^{-2} \, Hz^{-1}}$) according to $L_{X,UV} = 4 \, \pi \, d_l^2  \,F_{X,UV}$ where $d_l$ is the luminosity distance in cm and the K-correction is assumed to be equal to 1 for Quasars \citep{Lusso2020A&A...642A.150L}. Here we consider a flat $\Lambda$CDM model.
Our main gold Quasar sample of 983 sources is obtained by fixing $\Omega_M = 0.3$ and $H_0 = 70  \, \mathrm{km} \, \mathrm{s}^{-1} \, \mathrm{Mpc}^{-1}$ in the distance luminosity. The values of $k_{UV}$ and $k_X$ used to determine $L'_{UV}$ and $L'_X$ are $k_{UV} = 4.36 \pm 0.08$ and $k_X = 3.36 \pm 0.07$ \citep{DainottiQSO2022}, which have been obtained within a flat $\Lambda$CDM model with $\Omega_M = 0.3$ and $H_0 = 70  \, \mathrm{km} \, \mathrm{s}^{-1} \, \mathrm{Mpc}^{-1}$, consistently with our assumption. The evolutionary parameter $k$ depends on $\Omega_M$, but not on $H_0$ \citep{DainottiQSO2022,2023MNRAS.518.2201D,LenartbiasfreeQSO2022}. Thus, when we change the value of $\Omega_M$ to $\Omega_M = 0.1$ and $\Omega_M = 1$ to test the dependence of our results on the cosmological assumptions, we accordingly use the values of $k_{UV}$ and $k_X$ corresponding to these cosmologies. More precisely, these values are obtained from the functions $k_{UV} (\Omega_M)$ and $k_{X} (\Omega_M)$ reported in \cite{DainottiQSO2022} and shown in their Fig. 4. The resulting values are $k_{UV} = 4.79 \pm 0.08$ and $k_X = 3.81 \pm 0.06$ for $\Omega_M = 0.1$ and $k_{UV} = 3.89 \pm 0.08$ and $k_X = 2.88 \pm 0.06$ for $\Omega_M = 1$.

The choice of a specific $\Omega_M$ and thus, a value of $k$ for the correction for evolution automatically induces the circularity problem. This issue can be completely overcome in the cosmological fit if we do not fix $k$ a-priori to compute the luminosities, but we apply the functions $k_{UV} (\Omega_M)$ and $k_{X} (\Omega_M)$ in the fitting procedure while leaving also the cosmological parameters free to vary, as already done in \cite{LenartbiasfreeQSO2022} for Quasars and  in \cite{2023MNRAS.518.2201D} for GRBs. This allows us to avoid assuming a-priori an underlying cosmology and to leave the evolutionary parameters free to vary together with the parameters of the fit. In our work, we also employ this circularity-free methodology when we fit $\Omega_M$ with the gold sample of 975 sources selected by applying the $\sigma$-clipping procedure to the relation between X-ray and UV fluxes, instead of luminosity. Indeed, the selection of the sample based on measured fluxes does not require any assumption on the cosmological model, contrary to the one based on luminosities, and, thus enables us to apply this circularity-free procedure to fit $\Omega_M$ with a sample that is not biased by any cosmological assumption.

\subsection{Fitting procedure}

We have performed all fits with the Bayesian D'Agostini method \citep{2005physics..11182D}.
The likelihood function ($\mathcal{L}$) employed to constrain parameters with Quasars is given by \cite{2020MNRAS.492.4456K,2020MNRAS.497..263K,Lusso2020A&A...642A.150L,2021MNRAS.502.6140K,2022arXiv220611447C,2022MNRAS.510.2753K,2022arXiv220310558C,Bargiacchi2022MNRAS.515.1795B,LenartbiasfreeQSO2022} as:
\begin{equation} \label{lfqso}
\text{ln}\mathcal{L} = -\frac{1}{2} \sum_{i=1}^{N} \left[ \frac{(y_{i}-\phi_{i})^{2}}{s^{2}_{i}} + \text{ln}(s^{2}_{i})\right]
\end{equation}
where ``ln" is the natural logarithm and $N$ is the number of sources. When we fit the luminosities, $y_i=\mathrm{log_{10}} L'_{X}$ of the Quasar at redshift $z_i$, while $\phi_i$ is the logarithmic X-ray luminosity predicted from the fitted model. The quantity $s^{2}_{i} = (\Delta \mathrm{log_{10}} L'_{X})^{2}_{i} + \gamma^{2} (\Delta \mathrm{log_{10}} L'_{UV})^{2}_{i} + \delta^{2}$ includes the statistical 1 $\sigma$ uncertainties ($\Delta$) on both luminosities and the intrinsic dispersion $\delta$ of the RL relation. The free parameters of the fit are $\gamma$, $\beta$, $\delta$, and the ones of the cosmological model studied. 
The same methodology is applied when we fit fluxes instead of luminosities, just replacing $\mathrm{log_{10}} L'_{UV}$ and $\mathrm{log_{10}} L'_{X}$ with the measured $\mathrm{log_{10}} F_{UV}$ and $\mathrm{log_{10}} F_{X}$. In this case, $\gamma_\mathrm{F}$ and $\delta_\mathrm{F}$ are the slope and the intrinsic dispersion of the linear relation.

\subsection*{$\sigma$-clipping technique}

Before applying the $\sigma$-clipping procedure, we have searched for possible outliers. Thus,
we have computed the maximum value of $\Delta \mathrm{log_{10}} L'_{UV} / \mathrm{log_{10}} L'_{UV}$ and $\Delta \mathrm{log_{10}} L'_{X} / \mathrm{log_{10}} L'
_{X}$. Points with large uncertainties on luminosities would not be removed by the $\sigma$-clipping and would have a higher weight in the cosmological fits compared to points with smaller uncertainties. This check has revealed that our sample does not present any outliers of this kind. Independently of the $\Omega_M$ value, the maximum values of $\Delta \mathrm{log_{10}} L'_{UV} / \mathrm{log_{10}} L'_{UV}$ and $\Delta \mathrm{log_{10}} L'_{X} / \mathrm{log_{10}} L'_{X}$ are 0.022 and 0.013, respectively, with only one source with $\Delta \mathrm{log_{10}} L'_{UV} / \mathrm{log_{10}} L'_{UV} > 0.02$ and four sources with $\Delta \mathrm{log_{10}} L'_{UV} / \mathrm{log_{10}} L'_{UV} > 0.01$. Because these values are lower than $\sim 1-2 \%$, we do not remove any Quasar from the initial sample. Hence, we apply the $\sigma$-clipping selection to all 2421 sources.

The $\sigma$-clipping allows us to reduce the intrinsic scatter of the RL relation by removing the sources with a vertical distance from the best-fit relation greater than a chosen threshold value by assuming a given cosmological model. It is used when dealing with relations presenting an intrinsic dispersion to remove possible outliers in the sample and assure a better determination of the free parameters of the relation. This procedure has already been successfully applied for Quasars to constrain cosmological parameters \citep{Lusso2020A&A...642A.150L,2021A&A...649A..65B, Bargiacchi2022MNRAS.515.1795B}. We detail the method for our case. First, we fit the RL relation with the whole Quasar sample assuming a flat $\Lambda$CDM model with $H_0 = 70  \, \mathrm{km} \, \mathrm{s}^{-1} \, \mathrm{Mpc}^{-1}$ and $\Omega_M = 0.3$ or $\Omega_M = 0.1$ or $\Omega_M = 1$, to compute the X-ray and UV luminosities. The model used is the RL relation, thus $\phi_i$ of Equation \ref{lfqso} is $\phi_i = \gamma \, \mathrm{log_{10}} L'_{UV,i} + \beta$ for the source at redshift $z_i$. From this fit, we obtain the best-fit values of $\gamma$, $\beta$, and $\delta$ and their 1 $\sigma$ uncertainties. Then, we evaluate for each Quasar at $z_i$ the quantity
\begin{equation}
\label{sigmaclip}
\Sigma = \frac{|\mathrm{log_{10}} L'_{X,i} - (\gamma \, \mathrm{log_{10}} L'_{UV,i} + \beta)|}{\sqrt{\Delta^2 \mathrm{log_{10}} L'_{X,i} + \gamma^2 \, \Delta^2 \mathrm{log_{10}} L'_{UV,i} + \delta^2}}
\end{equation}
where we use the best-fit values previously obtained for $\gamma$, $\beta$, and $\delta$. This quantity is exactly the one that is minimized in the fitting algorithm (i.e. the first term in square bracket of Equation \ref{lfqso}) to determine the parameters of the RL relation, and thus it is the most appropriate to estimate the discrepancy between the measured X-ray luminosity and the one predicted from the RL relation. 
Once we have computed $\Sigma$ for each source, we select only Quasars with $\Sigma \leq  \sigma_{\mathrm{clipping}}$ and we repeat the fit of the RL relation with this reduced sample. Since the fit on this new sample yields best-fit values of $\gamma$, $\beta$, and $\delta$ different from the ones of the previous fit, and thus different $\Sigma$ values for each source, there will be Quasars in the sample considered at this step that do no fulfil the requirement $\Sigma \leq \sigma_{\mathrm{clipping}}$ anymore. Hence, we iterate this procedure until all sources in the selected sample verify the requisite. After this $\sigma$-clipping process, we obtain a final Quasar sample, with the corresponding best-fit values and 1 $\sigma$ uncertainties of $\gamma$, $\beta$, and $\delta$. We have chosen several $\sigma_{\mathrm{clipping}}$ values between 0.6 and 2 to investigate how the assumed $\sigma_{\mathrm{clipping}}$, the best-fit value of $\delta$, and the number of survived sources are related. 
This method selects the gold samples (983 Quasars) within a flat $\Lambda$CDM model shown in the left panel of Fig. \ref{fig:lxluv} with its corresponding best-fit RL relation (purple line).
We have also applied the same method to select the Quasar samples using the observed fluxes instead of the luminosities. 
We fit the relation between $\mathrm{log_{10}} F_{X}$ and $\mathrm{log_{10}} F_{UV}$ for which $\gamma_\mathrm{F}$, $\beta_\mathrm{F}$, and $\delta_\mathrm{F}$ are the slope, intercept, and intrinsic dispersion, respectively. The gold sample of 975 Quasars obtained with the $\sigma$-clipping on fluxes is shown in the left panel of Fig. \ref{fig:fxfuv} with the best-fit linear relation (purple line).

\subsection{Cosmological fits}

For each Quasar sample produced by the $\sigma$-clipping with a specific threshold $\sigma_{\mathrm{clipping}}$ we have fitted a flat $\Lambda$CDM model. Following the fitting procedure described above, if we use the Quasar sample selected through the $\sigma$-clipping on the luminosities corrected for a fixed redshift evolution, the quantities $y_i$ and $\phi_i$ of Equation \ref{lfqso} for a source at $z_i$ are respectively $y_i = \mathrm{log_{10}} F_{X,i} + \mathrm{log_{10}}(4 \, \pi \, d_l^2(z_i)) - k_X \mathrm{log_{10}} (1+ z_i)$ and $\phi_i = \gamma \, [ \mathrm{log_{10}} F_{UV,i} + \mathrm{log_{10}}(4 \, \pi \, d_l^2(z_i)) - k_{UV} \mathrm{log_{10}} (1+ z_i) ] + \beta$, where $k_X$ and $k_{UV}$ are the evolutions corresponding to the cosmological model assumed. Instead, if we perform the fit on the sample obtained from the $\sigma$-clipping on the measured fluxes, we require $y_i = \mathrm{log_{10}} F_{X,i} + \mathrm{log_{10}}(4 \, \pi \, d_l^2(z_i)) - k_X(\Omega_M) \, \mathrm{log_{10}} (1+ z_i)$ and $\phi_i = \gamma \, [ \mathrm{log_{10}} F_{UV,i} + \mathrm{log_{10}}(4 \, \pi \, d_l^2(z_i)) - k_{UV}(\Omega_M) \, \mathrm{log_{10}} (1+ z_i) ] + \beta$, thus avoiding the circularity problem. The luminosity distance $d_l$ is computed fixing $H_0 = 70  \, \mathrm{km} \, \mathrm{s}^{-1} \, \mathrm{Mpc}^{-1}$ and considering $\Omega_M$ as a free parameter with a wide uniform prior between 0 and 1. We also leave $\gamma$ and $\beta$ free to vary. Hence, we obtain the best-fit values of $\Omega_M$, $\gamma$, $\beta$, and $\delta$ with their associated 1 $\sigma$ uncertainty. 

\section{Results}\label{results}

\subsection{Golden samples with an assumed cosmology}\label{golden sample}
\begin{figure}[h!]
    \includegraphics[width=0.58\textwidth]{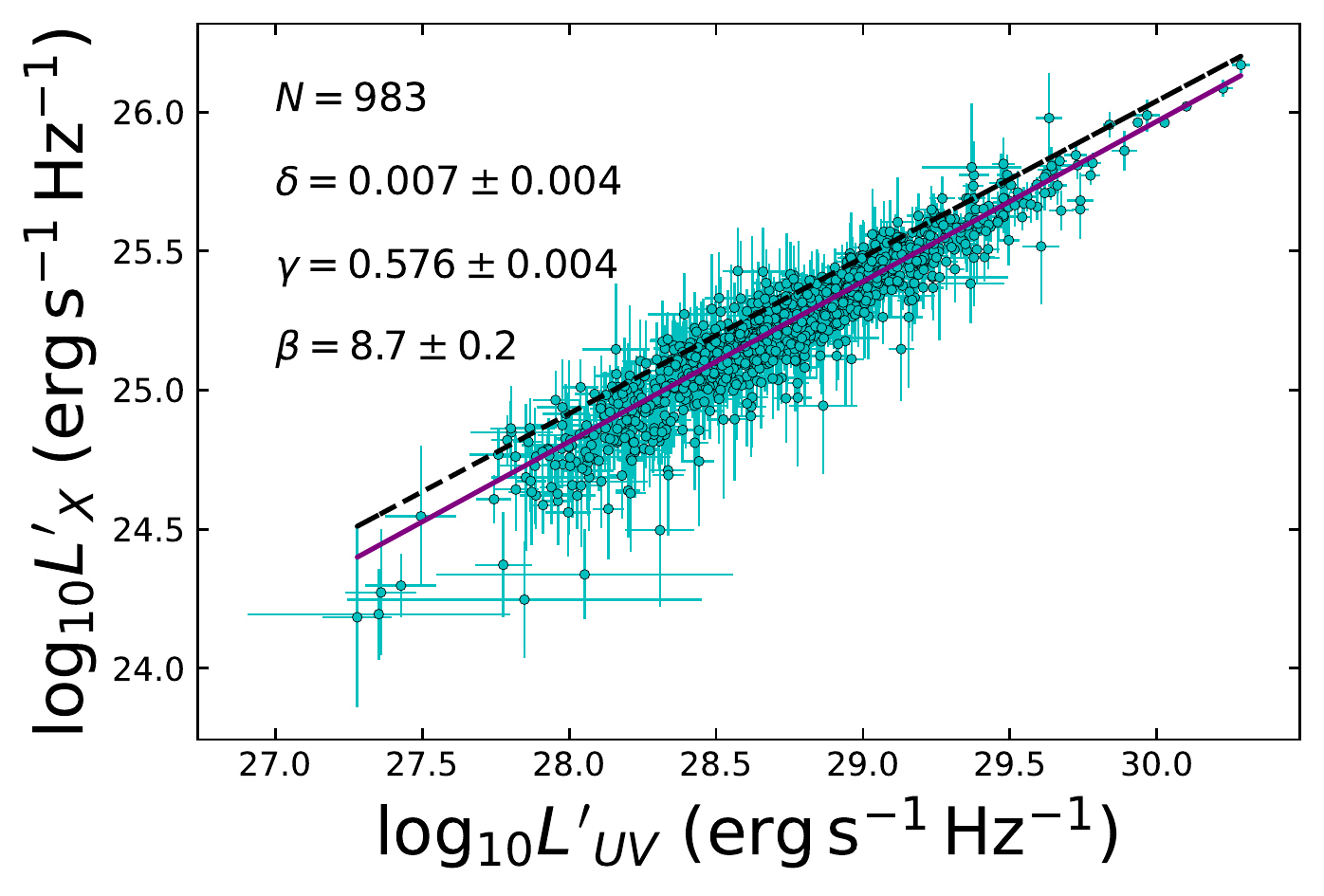}
    \includegraphics[width=0.40\textwidth]{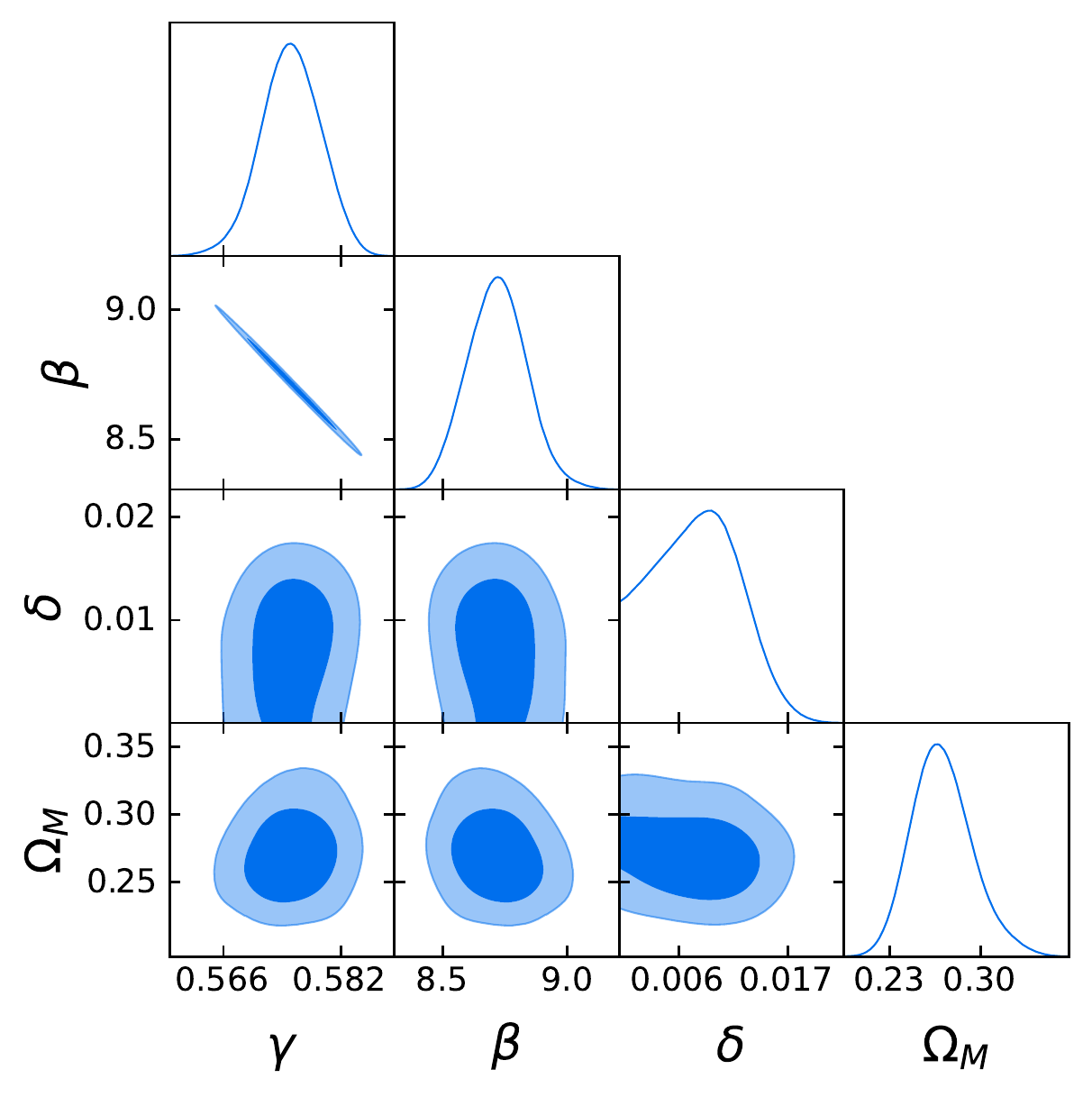}
    \caption{Left panel: The golden sample of 983 Quasars obtained with the $\sigma$-clipping on the $L_\mathrm{X} - L_{\mathrm{UV}}$ relation assuming a flat $\Lambda$CDM model with $\Omega_M= 0.3$ and $H_0 = 70  \, \mathrm{km} \, \mathrm{s}^{-1} \, \mathrm{Mpc}^{-1}$. The resulting best-fit values of the parameters are: the slope $\gamma = 0.576 \pm 0.004$, the normalization $\beta= 8.7  \pm 0.2$, and the dispersion $\delta = 0.007 \pm 0.004$. Blue points are the sources with error bars representing the statistical 1 $\sigma$ uncertainties and the best-fit linear relation is drawn as a purple line. The black dashed line is the best-fit line for the 2036 Quasars used in \citet{Lusso2020A&A...642A.150L}, for which $\gamma = 0.562 \pm 0.011$, $\beta= 9.2 \pm 0.3$, and $\delta = 0.221 \pm 0.004$ after correction for evolution.  Right panel: Cosmological results in a flat $\Lambda$CDM model from our golden sample shown in the left panel. This shows the values of $\Omega_M$, $\gamma$, $\beta$, and $\delta$. The contour levels at $68 \%$ and $95 \%$ are represented by the inner dark and light blue regions, respectively.} 
    \label{fig:lxluv}
\end{figure}

Our initial Quasar sample is the most up-to-date  for cosmological studies \citep{Lusso2020A&A...642A.150L} and composed of 2421 sources between $z=0.009$ and $z=7.54$.
This sample is studied to define the gold Quasar sample to obtain the tightest possible relation to be used as an efficient cosmological tool with the same precision achieved with SNe Ia.
Differently from recent works \citep{Lusso2020A&A...642A.150L,2021A&A...649A..65B,Bargiacchi2022MNRAS.515.1795B}, we use the full sample of Quasars at all redshifts and we correct for redshift evolution of the sample, as detailed before. So, we retain also sources at small redshifts instead excluded in most of previous analyses.
Differently from \cite{rl19}, the RL relation is already corrected for selection biases and redshift evolution as shown in \cite{DainottiQSO2022} when the $\sigma$-clipping is applied. This procedure allows us to obtain the smallest intrinsic scatter of an already bias-free and evolution-free relation. Nevertheless, to constrain cosmological parameters, such as $\Omega_M$, we are not only interested in reaching the smallest dispersion, but also in relying on a statistically sufficient number of sources. Thus, we need to find a compromise between these two antagonistic factors. Indeed, increasing the data set results in a larger intrinsic dispersion and vice versa.
We divide our results in the search of two main golden samples, one assuming a given cosmological model and the second one without any cosmological assumption, thus completely overcoming the so-called circularity problem.

If we assume a flat $\Lambda$CDM model with $\Omega_M = 0.3$ and $H_0 = 70  \, \mathrm{km} \, \mathrm{s}^{-1} \, \mathrm{Mpc}^{-1}$, the optimal number of sources is 983, obtained by requiring a threshold for the $\sigma$-clipping $\sigma_{\mathrm{clipping}} = 1.788$, with a scatter $\delta = 0.007$ of the RL relation. The 983 Quasars created in such a manner define the golden sample shown with the corresponding best-fit RL relation in the left panel of Fig. \ref{fig:lxluv}.


\begin{figure}[ht!]
  \centering
 \includegraphics[width=140mm]{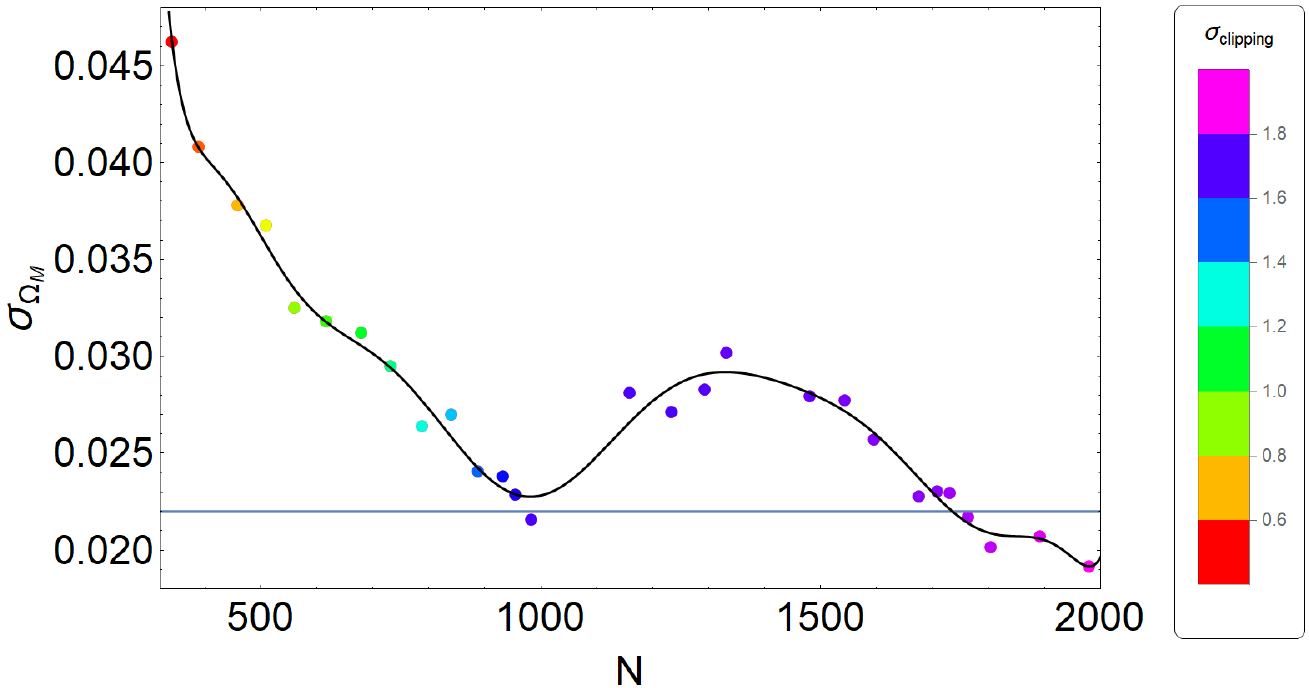}
 \includegraphics[width=140mm]{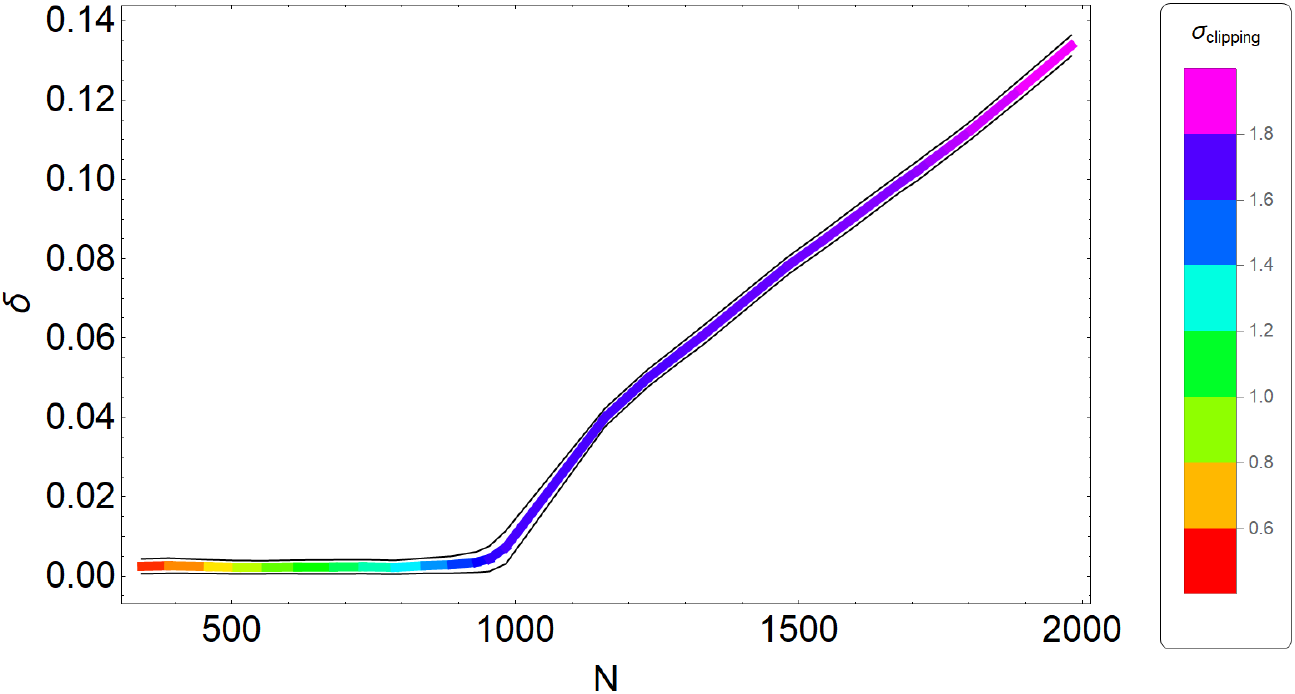}
 \caption{Upper panel: The uncertainty of $\Omega_{M}$ as a function of the number of sources and as function of the sigma-clipping showed with the colour bar on the right. The solid black line shows the best-fit of the decreasing trend of the uncertainty on $\Omega_{M}$ vs. the number of sources, while the horizontal blue line denotes the uncertainty on $\Omega_{M}$ reached with the Pantheon SNe Ia sample. This is obtained for our golden sample under the assumption of a flat $\Lambda$CDM model with $\Omega_M = 0.3$, and $H_0 = 70  \, \mathrm{km} \, \mathrm{s}^{-1} \, \mathrm{Mpc}^{-1}$. Lower panel: The intrinsic dispersion of the RL relation as a function of the number of sources and as a function of the sigma-clipping indicated on the right as a color bar. Black lines mark the 1 $\sigma$ uncertainty on the $\delta$ values.} 
   \label{fig:precision}
\end{figure}

\begin{figure}[ht!]
  \centering
        \includegraphics[width=138mm]{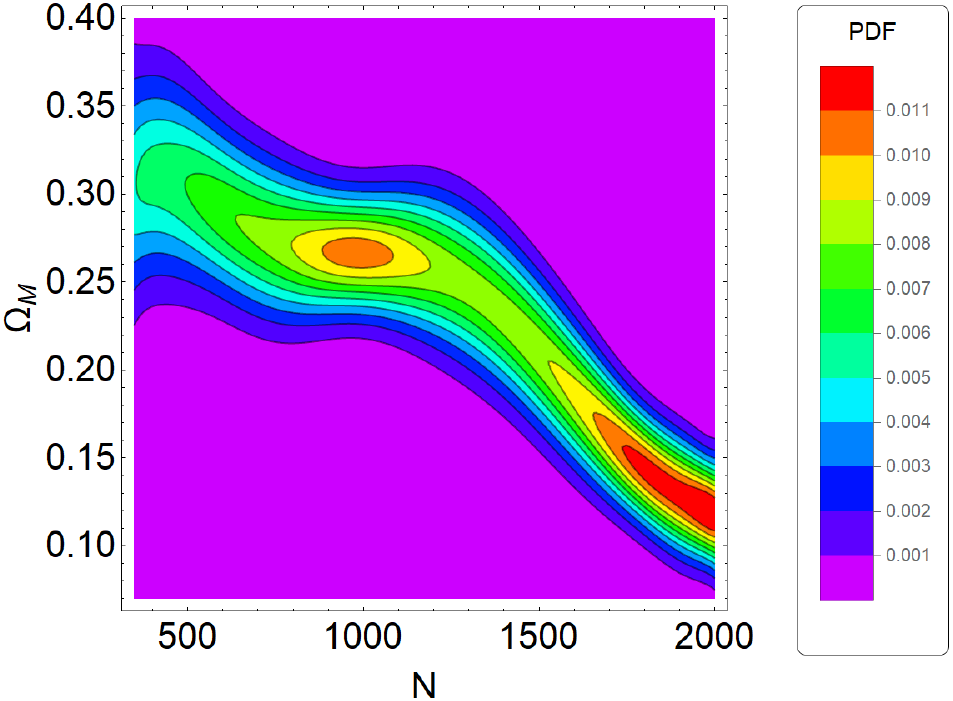}    
        \includegraphics[width=140mm]{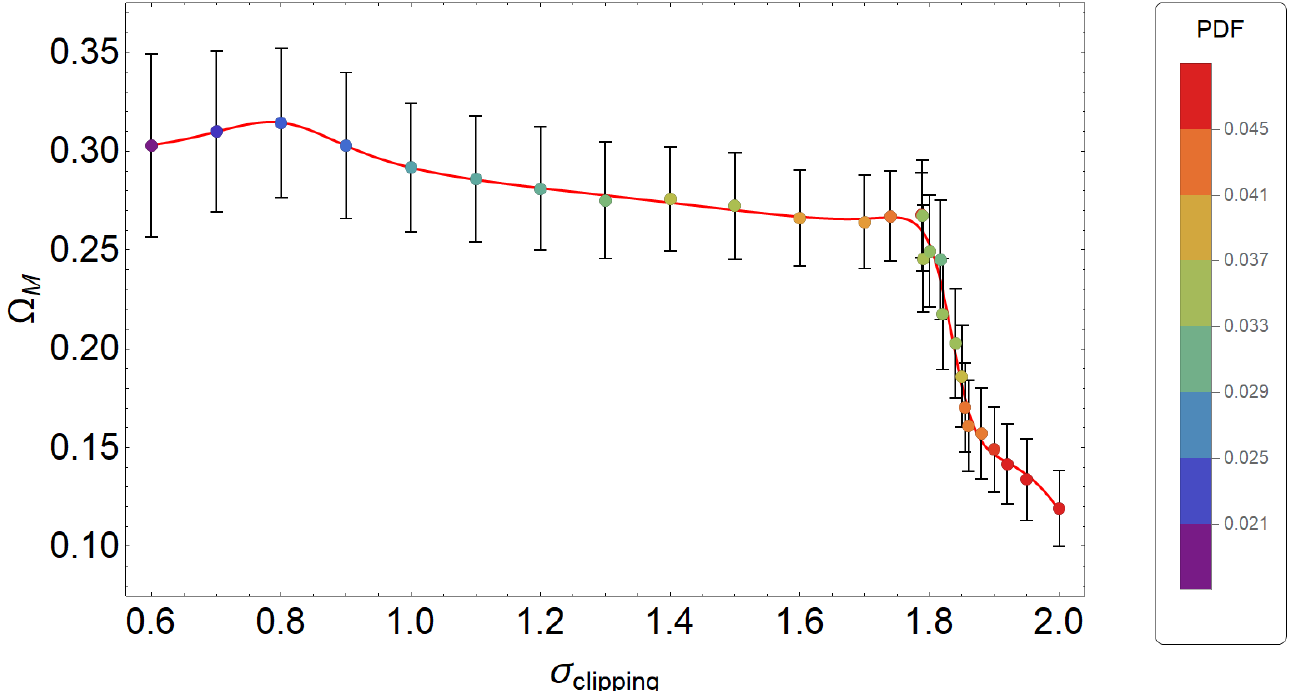} 
    \caption{Upper panel: The values of $\Omega_M$ and their associated uncertainties vs. the number of Quasars. The color bar on the right shows the normalized probability density, indicating for each sample size the most probable value of $\Omega_{M}$, thus the smallest uncertainty on $\Omega_{M}$. This Fig. indicates that the smallest error bar on $\Omega_{M}$ (the red contour) is achieved for $N\approx 2000$, which yields $\Omega_{M}=0.119\pm 0.019$. This is obtained for our golden sample assuming a flat $\Lambda$CDM model. Bottom panel: Values of $\Omega_M$ with corresponding 1 $\sigma$ uncertianities as a function of the $\sigma$-clipping threshold and the probability distribution function (PDF) showed with the colour bar on the right side. The red line is the best-fit of $\Omega_M$ points.} 
    \label{fig:probability}
\end{figure}

Then, we use this sample to derive $\Omega_M$ with the Monte Carlo Markov Chain (MCMC) computation and leaving contemporaneously free the values of $\Omega_M$ in the range from 0 a 1 using a uniform distribution and the parameters of the RL relation: $\gamma$ (the slope), $\beta$ (the normalization) and $\delta$ (the dispersion). We fix $H_0 = 70  \, \mathrm{km} \, \mathrm{s}^{-1} \, \mathrm{Mpc}^{-1}$ assuming a standard flat $\Lambda$CDM model. 
Results of this analysis are shown in the right panel of Fig. \ref{fig:lxluv} where the corner plot of the RL relation parameters and $\Omega_M$ are obtained when we perform the correction for the redshift evolution. The best-fit value yields $\Omega_M = 0.268 \pm 0.022$ which carries the same uncertainty on $\Omega_M$ found with the Pantheon sample (1048 SNe Ia) assuming the same cosmology \citep{scolnic2018}, as shown with the horizontal blue line in Fig. \ref{fig:precision}. 
We also recover the $\Omega_M$ we have assumed to build this sample within 0.68 $\sigma$. 

To assess to what extent the number of sources impacts the values and the uncertainty on $\Omega_M$, we show in the upper panel of Fig. \ref{fig:precision} the values of $\sigma_{\Omega_M}$ vs. the number of sources and the varying of the sigma-clipping shown as a color bar.
In the upper panel of Fig. \ref{fig:precision}, we start from the total sample of sources and we apply a very restrictive criterion for the $\sigma$-clipping=0.6 and then we continue with larger values of $\sigma$-clipping until arriving at 2.0. When we change the values of the $\sigma$-clipping, the smaller is the $\sigma$-clipping, the smaller the sample size obtained if we consider sources from 300 to 1000 and from 1700 to 2000, but when the $\sigma$-clipping is too small (0.045 for example, see the red point in the upper panel of Fig. \ref{fig:precision}), then the uncertainties on $\Omega_M$ becomes larger. 
Since the $\sigma$-clipping and the number of sources also determine the intrinsic dispersion of the RL, thus we show in the bottom panel of Fig. \ref{fig:precision} the intrinsic scatter as as a function of the number of sources and of the $\sigma$-clipping shown as a color bar.
It is clear from this figure that the dispersion of the RL relation increases monotonically at the increase of the sample sizes starting from around 1000 sources, while to achieve smaller dispersion the trend is rather flat from 300 to 1000 sources. 
This increasing trend of the dispersion of RL relation as a function of the number of sources is reflected by the increasing values of the uncertainties on $\Omega_M$ in the range between 1000 and 1300. However, the highly non-linear process of obtaining cosmological parameters, in this case the value of $\sigma_{\Omega_M}$, and the cut that the $\sigma$-clipping induces in the initial sample does not allow a straightforward comparison between the upper and lower panel of Fig. \ref{fig:precision}. 
To conclude, the number of 975 sources is the optimal compromise to obtain the smallest uncertainty on $\Omega_M$ taking into consideration the dispersion. Indeed, if we still would enlarge further the $\sigma$-clipping the number of sources would become larger together with the dispersion of the RL relation. This is the reason why we have mentioned before the relevance to reach a compromise among the number of sources used and the dispersion of the RL and consequently the uncertainty on $\Omega_M$ before the increase we observe between 1000 and 1700 sources. We stress that the subsamples shown in Fig. \ref{fig:precision} are not subsample of the Gold Sample, but they are samples drawn independently, however with the same procedure of the $\sigma$-clipping starting from the full sample assuming a flat $\Lambda$CDM model.
Continuing on the evaluation of how many number of sources are needed to obtain the most probable value of $\Omega_M$ we show  the colour map in the upper panel of Fig. \ref{fig:probability} where the probability density function (PDF) of $\Omega_M$ is plotted as a function of the number of sources.  
We note from this figure that 983 sources are the best sample as it provides closed cosmological contours with the highest corresponding probability. 
To complement this information, we also plotted in the bottom panel of Fig. \ref{fig:probability} the $\Omega_M$ values as a function of the $\sigma$-clipping and as a function of the probability density for $\Omega_M$ shown in the color bar. From this figure it is clear that the $\sigma$-clipping of 1.8 is the optimal value to choose our Gold sample since the scatter on $\Omega_M$ becomes small enough to reach the precision of SNe Ia Pantheon sample.
To check this result against the cosmological settings, we also test the assumptions of $\Omega_M=1$ (Universe filled by matter in which $\Lambda= 0$), and $\Omega_M=0.10$ (very close to the De Sitter Universe). We obtain that the best-fit values of $\Omega_M$ are consistent within less than 3 $\sigma$ with the a-priori assumption.
In fact, when we choose the golden sample of 980 Quasars derived with the $\sigma_{\mathrm{clipping}}=1.788$ assuming $\Omega_M=0.10$, we obtain $\Omega_M=0.083 \pm 0.009$, while when we select the sample of 968 Quasars assuming $\Omega_M=1$, we obtain $\Omega_M=0.910 \pm 0.055$ for $\sigma_{\mathrm{clipping}}=1.785$.

\subsection{The Anderson Darling Test for different Gold samples and the parent population}
To check the similarity between the Gold sample and the parent population, we have applied the Anderson-Darling test to compare both the fluxes-fluxes distributions and the luminosity-luminosity distributions from the parent population and the Gold sample. The result of the test shows that the Gold of the luminosities and fluxes are not compatible both in X-rays and UV with the parent population. 
In addition, the samples from the parent population used by \cite{Lusso2020A&A...642A.150L} also is not compatible with our Gold sample. 
It is not surprising that we can have different results when other probes are added to Quasars since the probes with smaller uncertainties ($s$ in Eq. \ref{lfqso}) weigh more than the ones with larger uncertainties, being the $s$ values at the denominator of the likelihood function.
In addition, since our Gold sample is already a sample for which the selection biases have been removed, the cosmological results from this sample should not necessarily be the same as the enlarged  sample (the parent population) if  the parent population undergoes selection biases and redshift evolution.
Indeed, results of cosmological parameters may change if evolutionary effects are not considered \citep{Dainotti2013a,Dainotti2022MNRAS.514.1828D,Dainotti2023MNRAS.518.2201D, LenartbiasfreeQSO2022}.
In relation instead of our Gold sample both in X-rays and UV derived from a given cosmology ($\Omega_M=0.3$ and $H_0=70 \, km s^{-1} Mpc^{-1}$) these are compatible with the Gold sample originated assuming other cosmologies ($\Omega_M=0.1$, $H_0=70 \, km s^{-1} Mpc^{-1}$, and $\Omega_M=1$ and $H_0=70 \, km s^{-1} Mpc^{-1}$).
This ensures that the selection of our Gold sample does not depend on cosmological models.

\subsection{A circularity-free golden sample}\label{circularity free}

\begin{figure}[ht!]
    \centering
    \includegraphics[width=0.59\textwidth]{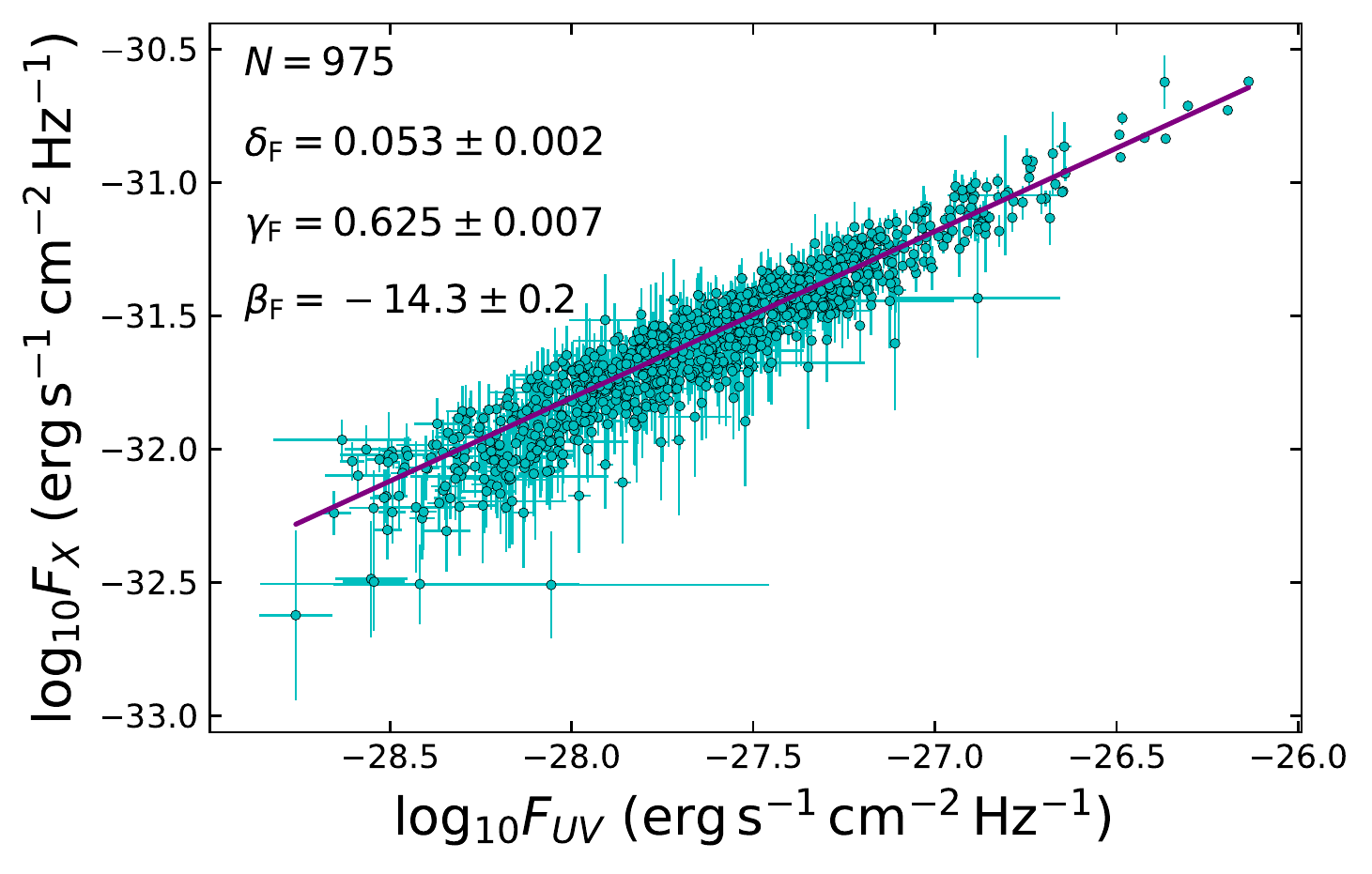}
     \includegraphics[width=0.39\textwidth]{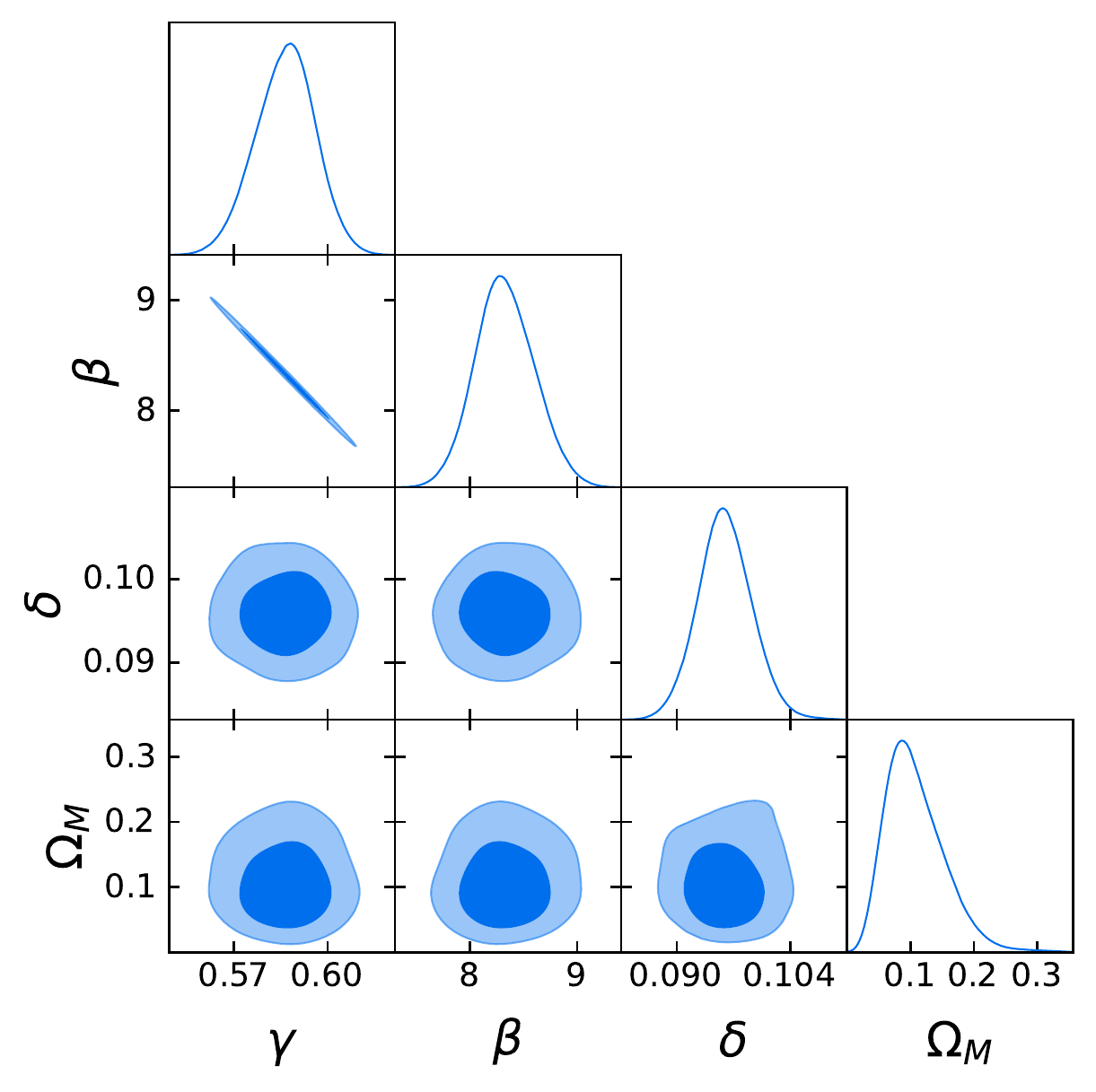}
    \caption{Left panel: 
    The gold sample of 975 Quasars generated with the $\sigma$-clipping on the $F_\mathrm{X} - F_{\mathrm{UV}}$ relation with the best-fit parameters being $\gamma_\mathrm{F} = 0.625 \pm 0.007 $, $\beta_\mathrm{F} = -14.3 \pm 0.2$, and $\delta_\mathrm{F} = 0.053 \pm 0.002$. Blue points are the sources with error bars representing the statistical 1 $\sigma$ uncertainties and the best-fit linear relation is drawn as a purple line. Right panel: Cosmological results from the golden sample shown in the left panel. This corner plot shows the values of $\Omega_M$, $\gamma$, $\beta$, and $\delta$. The contour levels at $68 \%$ and $95 \%$ are represented by the inner dark and light blue regions, respectively.} 
    \label{fig:fxfuv}
\end{figure}

To further guarantee that we completely avoid the circularity problem for the choice of the golden sample of Quasars, we use the $F_{X}-F_{UV}$ relation, the observer frame relation corresponding to the RL relation. As previously, we apply the same $\sigma_{clipping}$ procedure to reduce the scatter of the $F_{X}-F_{UV}$ relation. A $\sigma_{\mathrm{clipping}}=1.78$ identifies an optimal sample of 975 sources, shown in the left panel of Fig. \ref{fig:fxfuv}. 
We then use this sample (free from any circularity problem) to derive $\Omega_M$ and the RL parameters (see Methods), and we obtain $\Omega_M=0.107 \pm 0.047$, as reported in the right panel of Fig. \ref{fig:fxfuv}. \\

Additionally, to assure that our findings are not driven by the low-$z$ Quasars ($z<0.7$), which according to \cite{Lusso2020A&A...642A.150L} could be affected by host galaxy contamination and lower data quality, we have removed these sources (47 Quasars) from the golden sample obtaining $\Omega_M= 0.125 \pm 0.040$.

\section{Monte Carlo Markov Chain (MCMC) Sampling Uncertainty}\label{MCMC}
To further guarantee that our results are not due to the use of a single run in the MCMC calculation and that the sampling procedure is stable, we show that the results of the computation when it is run 100 times. With this procedure, we obtain $<\Omega_{M}>=0.112  \pm 0.048$, where the symbol $<>$ denotes the average value.
We have investigated the reliability of the best-fit values and 1 $\sigma$ uncertainty on $\Omega_M$ obtained in each of the cosmological fits. Our results are derived by fitting the free parameters of the models studied with only one MCMC run. We test these results against the sampling error on the parameters derived in the sampling procedure. To this end, we have looped all the MCMC samplings 100 times for each model, then computing the mean values of $\Omega_M$ and its uncertainty. The results obtained with this method for both quantities and all the cosmological cases investigated in our analysis are shown in Table \ref{tab:my_label}. 
These results are completely consistent with the ones obtained from only one run of the MCMC, with a maximum discrepancy of 0.08 $\sigma$. 

\begin{table}[ht!]
    \centering
    \begin{tabular}{|c|c|c|c|c|c|}
    \hline
        $\sigma_{\mathrm{clipping}}$ & $\Omega_{\mathrm{M,start}}$ & $<\Omega_{M}>$ & $S_{<\Omega_{M}>}$ & $<\sigma_{\Omega_{M}}>$ & $S_{<\sigma_{\Omega_{M}}>}$ \\\hline\hline
        \multicolumn{6}{|c|}{Luminosities} \\\hline
        1.788 & 0.3 & 0.2681 & 0.0010 & 0.0223 & 0.0007 \\\hline
         1.788 & 0.1 & 0.0836 & 0.0004 & 0.0086 & 0.0004 \\\hline
        1.785 & 1 & 0.9141 & 0.0024 & 0.0536 & 0.0013 \\\hline\hline
        \multicolumn{6}{|c|}{Fluxes} \\\hline
        1.78 & - & 0.1124 & 0.0021 & 0.0475 & 0.0021 \\\hline
    \end{tabular}
    \caption{The results of the looped computations for 100 iterations. $S$ denotes the standard deviation of a given parameter, $\Omega_{\mathrm{M,start}}$ is the value of $\Omega_{M}$ assumed to obtain the corresponding golden sample with the threshold value $\sigma_{\mathrm{clipping}}$. The cases in which the $\sigma$-clipping is applied to the relation in luminosities or in fluxes are also shown separately.}
    \label{tab:my_label}
\end{table}

\section{Discussion and Conclusions}\label{discussions}
In the choice of the golden samples with a fixed cosmology we have first assumed a flat $\Lambda$CDM model with $\Omega_M= 0.3$ and $H_0 = 70  \, \mathrm{km} \, \mathrm{s}^{-1} \, \mathrm{Mpc}^{-1}$. 
This is a necessary starting cosmological model and value for $\Omega_M$, because our aim is to compare the results of our uncertainties with the ones obtained using the Pantheon sample for the same cosmological model. Regarding the sample size, we note that the 1048 Pantheon SNe Ia have been slimmed down from an original sample of 3473 events, with a cutting of the $70\%$ of the starting data set \citep{scolnic2018}. Instead, in our work, we reduce the initial sample of 2421 Quasars of the $\sim 60\%$ to build all the golden samples studied.
Thus, the slimming of our sample is even less severe than the one of SNe Ia.
We here compare the SNe Ia and Quasars sample from a mere statistical point of view and, not for the purpose to compare them from a physical and an observational point of view, since they differ on both aspects.

Our results highlight that the uncertainties on $\Omega_M$ depend on the assumed cosmology, namely the smaller the uncertainties the more we are closer to the most likely value of the cosmological parameters. The uncertainty on $\Omega_M$ assuming $\Omega_M=0.10$ is 6.7 times smaller than the one in the case of $\Omega_M=1$. 
With the evolutionary parameters included and with no information about the underlying cosmology, we still obtain a value of $\Omega_M$ compatible with the De-Sitter Universe at the 2.3 $\sigma$ level, and is not compatible with the current value of $\Omega_M=0.338 \pm 0.018$ found in \cite{Brout2022ApJ...938..110B} at the level of 4.58 $\sigma$ level. 
We here clarify that this estimate on $\Omega_M$ found by us is larger (more than 1 $\sigma$) than the observed baryonic matter today if we consider the value found by Planck measurements of CMB \citep{planck2018} which is $0.0500 \pm 0.0002$.  
Although the value of the uncertainties in the gold sample identified by the $F_{X}-F_{UV}$ relation is 2.14 times larger than the ones obtained by the Pantheon sample of SNe Ia \citep{scolnic2018}, we can still obtain uncertainties comparable with the 740 SNe Ia shown in \cite{Betoule2014A&A...568A..22B} where $\sigma_{\Omega_M} = 0.042$.

The results on $\Omega_M$ obtained with the golden sample of 975 quasars from the $F_{X}-F_{UV}$ relation (i.e. $\Omega_M= 0.107 \pm 0.047$) is compatible with $\Omega_M =0.125 \pm 0.040$ derived from the same golden sample to which we have removed the low-$z$ sources due to the contamination of the host galaxies (47 Quasars).
However, when a larger sample is available is necessary to check if the current results still hold within 1 $\sigma$.
Although there have been several studies that have measured $\Omega_M$ with Quasars with lower precision (e.g. \citealt{2020MNRAS.497..263K,2022MNRAS.510.2753K,2022arXiv220310558C}), we here for the first time obtain a value with a higher precision with QSOs alone, which is not due to a circular argument, since it is based on a flux-flux relation. In addition, the analysis performed by us using the luminosities and assuming a given cosmological model is meant to show the great potentiality of Quasars to be used as standardizable candles even currently when an appropriate sample size and reduced uncertainties is used. Indeed, the analysis we have shown here is similar to the analysis we performed in \citet{2023MNRAS.518.2201D} where we were not interested in knowing the value of the cosmological parameters, but we were focused on how many sources, in this case Quasars, are needed to reach the same precision of SNe Ia Pantheon sample.

In conclusion, we have shown that Quasars alone with the RL relation can now be upgraded to reliable standard candles to measure cosmological parameters such as $\Omega_M$ with the same precision as SNe Ia, but at a large redshift, up to 7.5, when a golden sample of Quasars is chosen.

\clearpage





\section{Acknowledgement}
We thank Beta Lusso and Risaliti Guido for the discussion on the role of selection biases in the sample and to Biagio De Simone to help running a couple of notebooks for the MCMC sampling.
\bibliographystyle{aasjournal}
\bibliography{bibliografia}



\end{document}